\documentclass{appolb}
\usepackage{epsfig}

\begin{document}
\title{Charmonium production at the LHC%
\thanks{Presented at Three Days of Strong Interactions, 9-11/07/2009 Wroc{\l}aw, Poland}%
}
\author{Magdalena Malek
\address{Institut de Physique Nucl\'{e}aire d'Orsay (IPNO) - France\\
CNRS: UMR8608 - IN2P3 - Universit\'{e} Paris Sud - Paris XI}
}
\maketitle
\begin{abstract}
We summarize the perspectives on quarkonium, in particular charmonium, detection at the LHC, both
for proton-proton and heavy-ion collisions. We give a review of the experimental capabilities of
the four LHC detectors: ALICE, ATLAS, CMS and LHCb.
\end{abstract}

\section{Introduction}
The measurement of the charmonium (J/$\psi$, $\psi^{'}$) resonances in nucleus-nucleus collisions
is considered to be one of the most powerful methods to probe the nature of the high density QCD
matter. First, the in-medium behavior of quarkonium is supposed to be one of the most direct
probes of the Quark Gluon Plasma (QGP) formation~\cite{satz}. It was predicted that in a
deconfined medium, the Debye screening of the colour potential leads to the melting of quarkonium
states and thus provides an estimate of the medium temperature. Second, at LHC energies heavy
quarks are produced mainly through gluon-gluon fusion process thus they are significantly affected
by parton dynamics in the low-$x$ regime. Therefore, they can be used as a tool to probe the
saturation of the gluon
density at low-$x$ in the nucleus and the Color Glass Condensate formation~\cite{CGCreviews}.\\
The LHC (Large Hadron Collider) will provide, at nominal operating conditions, Pb+Pb collisions at
$\sqrt{s_{NN}}$~=~5.5 TeV allowing the study of the quarkonium production in an unprecedented
energy regime. The LHC will deliver p+p and p+Pb collisions at $\sqrt{s_{NN}}$~=~14 and 8.8~TeV
respectively providing a solid baseline for the Pb+Pb system.

\section{From SPS to RHIC}
\subsection{Normal and anomalous suppression at SPS}
The J/$\psi$ yields (normalized to the Drell-Yan process) were measured at the CERN SPS for
different colliding systems and energies: from p+p to various p+nucleus (p+A) with proton energy
of 450~GeV. Moreover, nucleus-nucleus (A+A) collisions, from O+U and S+U (NA38) with a projectile
energy of 200~GeV per nucleon to Pb+Pb and In+In (NA50 and NA60) with a projectile energy of
158~GeV per nucleon have been studied.\\
It was observed that p+p, p+A and peripheral A+A data show an exponential suppression of the
J/$\psi$ to Drell-Yan ratio when plotted as a function of the average length of nuclear matter
traversed by the $c\overline{c}$ pair. This behavior was interpreted in terms of the normal
nuclear absorption (normal suppression) of the $c\overline{c}$ pair prior to the J/$\psi$
formation with a corresponding absorption cross section of 4.18~$\pm$~0.35~mb~\cite{abs}. For more
central A+A  collisions (In+In and Pb+Pb) an additional suppression (anomalous
suppression) was seen~\cite{addsupp}. This result was interpreted as a consequence of the creation of the deconfined medium.\\
One can thus think that the J/$\psi$ behaves as a predicted golden signature of the QGP formation.
\subsection{The RHIC anomalies}
The PHENIX experiment at the Relativistic Heavy Ion Collider (RHIC) measured the J/$\psi$
production in A+A (Au+Au) collisions at $\sqrt{s_{NN}}$~=~200~GeV. The J/$\psi$ behavior was
studied via the nuclear modification factor $R_{AA}$ defined as following:
$$R_{AA}~=~\frac{d^{2}N^{AA}_{J/\psi}/dp_{T}dy}{N_{coll}d^{2}N^{pp}_{J/\psi}/dp_{T}dy}$$
where $d^{2}N^{AA}_{J/\psi}/dp_{T}dy$ is the J/$\psi$ yield in Au+Au collisions, $N_{coll}$ the
corresponding mean number of binary collisions and $d^{2}N^{pp}_{J/\psi}/dp_{T}dy$ the yield in
p+p collisions. The PHENIX results brought up two surprises~\cite{phenix}:
\begin{itemize}
\item if plotted $R_{AA}$ as a function of the number of participants $N_{part}$:
at midrapidity the amount of suppression at RHIC is similar to the suppression observed at SPS.
$J/\psi$ were expected, as predicted in the color screening mechanism, to be more suppressed at
RHIC because of the higher energy density reached at collider energies.
\item if plotted $R_{AA}$ as a function of the number of participants $N_{part}$ for different
rapidity domains: $J/\psi$ are more suppressed at forward rapidity where the energy density is
lower than at midrapidity.
\end{itemize}



~\\
\indent Two possible reasons for theses results are explored and discussed. Firstly, the abundant
production of $c$ and $\overline{c}$ quarks at midrapidity domain can lead to their
recombination~\cite{recomb} and $J/\psi$ creation from initially uncorrelated quarks. Secondly,
the nuclear effects like standard gluon shadowing that are important at forward rapidity domain
have to be taken into account. Also the saturation effects (Color Glass
Condensate) could lead to the observed J/$\psi$ behavior.\\
For the moment, both interpretations are possible. There is not enough experimental data to
exclude one of them and to draw a definite conclusion.
\section{LHC}
The LHC collider will open up new perspectives in the study of the quarkonium of QCD. One can ask
what is different at the LHC in comparison with RHIC or SPS. First of all, an unprecedented
collision energy will be reached ($\sim$28 times bigger than at RHIC). Secondly, the LHC will
allow to access an unexplored Bjorken-$x$ region and thus probe the low-$x$ QCD phenomena
($x=\frac{M_{Q\overline{Q}}}{\sqrt{s_{NN}}} \exp(\pm y_{Q\overline{Q}})$). Finally, the expected
copious production of heavy quarks should help us to bring down the curtain on the different
predicted J/$\psi$ behaviors (suppression, regeneration or both simultaneously).\\
\indent There are several experimental challenges for quarkonium measurements at the LHC: a) the
choice of the reference production process for the normalization; b) estimation of the J/$\psi$
contribution originated from B-meson decays; c) understanding the complex combinatorial background
very well and d) taking into account possible heavy quark energy loss.\\These topics are under
investigation as they are crucial points needed to understand quarkonium production at the LHC.

\subsection{ALICE}
ALICE (A Large Ion Collider Experiment)~\cite{PPR1, PPR2, ALICE} experiment is dedicated to the
study of heavy ion collisions. Its goal is to study the properties of deconfined matter: the Quark
Gluon Plasma (QGP). The ALICE detector is composed of a central barrel system ($|\eta|<$0.9), a
muon spectrometer (-4.0$<\eta<$-2.5) and several small additional detectors. Quarkonium will be
measured in ALICE via two channels: the dielectron channel at
midrapidity and the dimuon channel at forward rapidity.\\
\indent The measurement via dielectron channel is provided by the combination of several detectors
that are here described, as seen by a particle travelling out from the interaction point:
\begin{itemize}
\item Inner Tracking System (ITS)~\cite{TDR_ITS} allows the reconstruction of the primary vertex,
secondary vertex finding, and particle identification via $dE/dx$. This detector is composed of
three subsystems of two layers each: the Silicon Pixel Detector, the Silicon Drift Detector and
the Silicon Strip Detector.
\item Time Projection Chamber (TPC)~\cite{TDR_TPC} allows track finding, momentum measurement,
and charged hadron identification via $dE/dx$. TPC has an inner radius of 0.9 m, an outer radius
of 2.5 m and a length of 5.1 m. The momentum resolution for the track reconstruction, including
TPC and ITS information, is expected to be better than 2\% for p$_{t}$~$<$~20~GeV/c.
\item Transition Radiation Detector (TRD)~\cite{TDR_TRD} allows electron identification and also provides
fast triggering. Electron identification is provided by the TRD for momenta larger than 1~GeV/c.
This detector is made of 18 longitudinal supermodules, 6 radial layers, and 5 stacks along the
beam axis.
\end{itemize}

The invariant mass resolution for the quarkonium was studied in the full simulation for Pb+Pb
collisions. The reconstructed peaks were fitted by a Gaussian and the invariant mass resolution
for the J/$\psi$ is found to be 30~MeV/c$^{2}$. The analysis showed that the J/$\psi$ signal can
be reconstructed up to p$_{t}$~=~10~GeV/c.


~\\
\indent The detection via dimuon channel in the forward rapidity region is provided by the muon
spectrometer~\cite{TDR_MUON}. This detector is composed of a set of absorbers: a front absorber, a
muon filter, a beam shielding and an absorber against the LHC background. The goal of these
absorbers is to suppress hadron and electron background. The tracking system allowing the muon
trajectory reconstruction contains five tracking stations. The dipole magnet with a field integral
of 3 Tm along the beam axis provides the bending power to measure the momenta of the muons. Two
trigger stations provide a fast electronic signal for the trigger selection of muon events. The
p$_{t}$ cut of 1~GeV/c applied to single muons allows charmonia detection down to zero transverse
momenta. The invariant mass resolution for the J/$\psi$ in Pb+Pb collisions is expected to be
around 70~MeV/c$^{2}$.
The high-p$_{t}$ reach for the J/$\psi$ is expected to be around 20~GeV/c.\\
~\\
\indent In addition to direct J/$\psi$ also those from B decays (secondary J/$\psi$) have to be
taken into account. The contribution of secondary J/$\psi$ to the total J/$\psi$ yields is about
20~\%. To separate the prompt J/$\psi$'s from the secondary ones, the measure of dielectron pairs
with displaced vertex must be performed. In fact, the secondary J/$\psi$ are produced at large
distances (several hundred of microns) from the primary vertex. The ITS vertexing capabilities
allow to perform this study in the central barrel. This analysis is not possible in the forward
rapidity region because the muon spectrometer does not provide vertexing capabilities.\\
~\\
\indent The suppression of J/$\psi$ was studied in the dimuon channel. Two extreme suppression
scenarios were considered. The first one characterized by a high critical deconfinement
temperature at T$_{c}$~=~270~MeV~\cite{SUPP_h} and the second one using
T$_{c}$~=~190~MeV~\cite{SUPP_l}. To check the detector capability for distinguishing between
different suppression scenarios the ratios of the resonance rates over those for beauty as a
function of the number of participants were studied. It was seen that the error bars for the
J/$\psi$ to open beauty
ratio are small enough to distinguish between these two suppression scenarios.\\
~\\
\indent To distinguish between different quarkonia production mechanisms the quarkonia
polarization study must be performed. The angular distribution of the decay products (dielectrons
or dimuons) in the quarkonia rest frame allows to reconstruct the polarization. It has been
predicted that the QGP formation may change the quarkonia polarization. In fact an increase of
J/$\psi$ polarization is expected in this case~\cite{Polar}.

\subsection{ATLAS}
The primary physics goal of the ATLAS experiment~\cite{ATLASTDR} is the search of the Higgs boson
and SUSY particles (supersymmetry). Nevertheless, other physics sectors like CP violation and rare
B-decays, can be explored. ATLAS will be able to provide evidence of physics phenomena beyond the
Standard Model (SM) in p+p collisions at $\sqrt{s_{NN}}$~=~14~TeV.\\
The ATLAS detector consists of four major components: inner tracking system ($|\eta|<$2.5),
calorimeters (electromagnetic: $|\eta|<$3.2, hadronic barrel: $|\eta|<$1.7, hadronic end-cap:
1.5$<|\eta|<$3.2 and forward: 3.1$<|\eta|<$4.9), muon spectrometer ($|\eta|<$2.5) and forward
detectors (LUCID: 5.3$<|\eta|<$6, ZDC: $|\eta|<$8.3 and ALFA).\\
ATLAS can detect the J/$\psi$ in the dimuon channel using the muon spectrometer covering
$|\eta|<$2.7 and full azimuth. The track coordinates measurement is provided by Monitored Drift
Tubes (barrel and end-cap regions) and by Cathode Strip Chambers (at large pseudorapidities and
close to the interaction point). The trigger system covers the pseudorapidity range $|\eta|\leq$
2.4. Resistive Plate Chambers (RPCs) are used in the barrel and Thin Gap Chambers (TGCs) in the
end-cap regions. The magnetic bending is provided by the large barrel toroid with a magnetic field
of 0.5~-~1~T. The strong magnetic field and the large material budget of the inner detector and
calorimeters make difficult the reconstruction of the muons with a $p_{T}~<$~3~GeV/c. The minimum
dimuon trigger threshold is around 4~GeV/c. The invariant mass resolution for J/$\psi$ is expected
to be around 54~MeV/c$^{2}$ and 68~MeV/c$^{2}$ respectively in p+p and Pb+Pb collisions.
\subsection{CMS}
The CMS experiment~\cite{CMSTDR, CMSHI} is dedicated to explore physics at the TeV scale. The
prime goals of CMS are to study the mechanism of electroweak symmetry breaking and provide
evidence of physics beyond SM. CMS will also study a large amount of signatures arising in the SM:
QCD, B-physics, diffraction, top quark properties, and electroweak physics topics such as the W
and Z boson
properties.\\
The CMS detector is composed of: inner tracking system ($|\eta|<$2.5), calorimeters
(electromagnetic: $|\eta|<$3, hadronic: $|\eta|<$5 ), muon system and few forwards detectors
(CASTOR: 5$<|\eta|<$6.6 and ZDC: $|\eta|>$8.3). In this experiment, the J/$\psi$ are detected via
their $\mu^{+}\mu^{-}$ decay by the muon system ($|\eta|<$2.4) that is divided into a central part
(Barrel Detector: $|\eta|<$1.2) and forward parts (Endcap Detector: $|\eta|<$2.4). The barrel
detector comprises the Drift Tubes chambers and Resistive Plates Chambers. The forward parts are
composed of a set of Cathode Strip Chambers in the 2 muon endcaps and a layers of Resistive Plates
Chambers. The CMS superconducting solenoid has a large bending power and can generate the magnetic
fields up to 4~T. The minimum muon momenta of 3.5 and 4~GeV/c respectively for barrel and endcap
region. The non-prompt fraction is estimated using the traditional method of the cut on pseudo
proper decay length. The high-p$_{t}$ reach for the J/$\psi$ is expected to be around 40~GeV/c.
The invariant mass resolution for the J/$\psi$ is
35~MeV/c$^{2}$ in Pb+Pb collisions. The CMS will also perform the polarization study of the quarkonium states.\\

\subsection{LHCb}
The main goal of the LHCb~\cite{LHCb} experiment is to study CP violation in the B meson systems
and to search for rare B decays. The LHCb apparatus is a single arm forward spectrometer with a
polar angular coverage from 15 to 300 mrad in the horizontal and 250 mrad in the vertical plane.
This corresponds to a pseudorapidity range 2$<\eta<$5. The choice of the detector geometry is
motivated by the fact that at high energies B hadrons are predominately produced in the forward
region. This detector is composed of a vertex detector system (VELO), a tracking system, aerogel
and gas RICH counters (Cherenkov detectors), an electromagnetic calorimeter with preshower
detector, a hadron
calorimeter and a muon detector.\\
The measurement of quarkonium will be performed in p+p collisions via J/$\psi
\rightarrow\mu^{+}\mu^{-}$. In fact, J/$\psi$ candidates are reconstructed by combining pairs of
oppositely charged tracks that originate from a common vertex. The muon trigger system requires a
muon with a p$_{t}~>$~1~GeV/c and one of the tracks with p$_{t}~>$~1.5~GeV/c. Both tracks have to
be identified as muons. In addition, one of the tracks is needed to be identified by cutting hard
on likelihood of the muon hypothesis relative to the
pion~\cite{mupion}. The expected mass resolution is around 11~MeV/c$^{2}$.\\
The LHCb detector will be able to separate prompt J/$\psi$'s from those from $b$ decays using the
$t$ variable defined as following:
$$ t~=~\frac{dz}{p_{z}^{J/\psi}}~\times~m^{J/\psi}$$
This variable is an approximation of the proper time of the $b$ quark in the forward region. This
distribution contains four components: a prompt component due to the J/$\psi$ originating from the
primary vertex (Gaussian distribution), an exponential component due to the J/$\psi$ coming from
$b$ decays, a combinatorial component due to particles coming from the primary vertex and a long
tail component due
to the association of the J/$\psi$ to the wrong primary vertex.\\
Moreover, the LHCb will measure the feed down from decays of $\chi_{c1, c2}$ to the prompt
J/$\psi$ and the J/$\psi$ polarization.

\section{Conclusions}
We presented an overview of perspectives of the four LHC experiments for quarkonium (charmonium)
physics. To understand the quarkonium production picture different colliding systems have to be
studied. The study of the p+p system allows to test the pQCD calculations and learn more about the
quarkonium production mechanisms. Additionally, the p+p system is a reference for more complicated
systems like p+Pb and Pb+Pb. The investigating of the p+Pb collisions allow us to evaluate the so
called Cold Nuclear Matter (shadowing, saturation, hadronic absorption) effects that can be very
important at the LHC energies. Finally, the study of the J/$\psi$ production in Pb+Pb collisions
allows to probe the properties of the QGP formation and evolution. The LHC will deliver p+p, p+Pb
and Pb+Pb collisions offering a new possibilities for the quarkonium study. The four LHC detectors
(ALICE, ATLAS, CMS and LHCb) have excellent experimental capabilities for quarkonium detection.
The large statistics of quarkonium expected at the LHC and performances of the LHC experiments
will help to clarify the current knowledge of this topic.
\section{Acknowledgments}
The author would like to thank Mercedes L\'{o}pez Noriega and David d'Enterria for reading of the
manuscrit and for constructive criticisms and suggestions.

\end{document}